\documentclass[10pt]{article}
\usepackage[a4paper, total={6in, 9in}]{geometry}

\usepackage[T1]{fontenc}
\usepackage{amsmath}
\usepackage{amssymb}
\usepackage{hyperref}
\usepackage{bm}
\usepackage{bbm}
\usepackage{tikz}
\usepackage{academicons}
\usepackage{graphicx}
\usepackage{xcolor}
\usepackage{booktabs}
\usepackage{subfig}
\DeclareMathOperator*{\argmax}{arg\,max}

\usepackage{arydshln}
\usepackage{float}
\usepackage{cite}

\begin{document}
\title{Deep Reinforcement Learning for Robust Goal-Based Wealth Management}
\author{Tessa Bauman\footnote{Corresponding author: \href{mailto:tessa.bauman@fer.hr}{tessa.bauman@fer.hr}}, Bruno Gašperov, Stjepan Begušić, and Zvonko Kostanjčar}

\date{\normalsize University of Zagreb, Faculty of Electrical Engineering and Computing,\\
Laboratory for Financial and Risk Analytics (\href{https://lafra.fer.hr}{lafra.fer.hr}), \\
Unska 3, 10000 Zagreb, Croatia}

\maketitle              
\begin{abstract}
Goal-based investing is an approach to wealth management that prioritizes achieving specific financial goals. It is naturally formulated as a sequential decision-making problem as it requires choosing the appropriate investment until a goal is achieved. Consequently, reinforcement learning, a machine learning technique appropriate for sequential decision-making, offers a promising path for optimizing these investment strategies. In this paper, a novel approach for robust goal-based wealth management based on deep reinforcement learning is proposed. The experimental results indicate its superiority over several goal-based wealth management benchmarks on both simulated and historical market data. 
\end{abstract}
\section{Introduction}
Goal-based wealth management (GBWM), also known as goal-based investing \cite{nevins}, is a relatively new class of approaches to wealth management that focus on attaining specific financial objectives (goals). As opposed to more traditional approaches to wealth management, in which the notion of expected profit and loss (PnL) plays a central role, GBWM revolves around maximizing the probability of goal attainment. Common investment goals include saving for college tuition, retirement, or purchasing a home. Recent years have seen an uptick in the popularity of GBWM~\cite{ref_article1}, particularly through the use of target date funds (TDFs). TDFs, also known as life-cycle funds \cite{gomes} or target-retirement funds, are mutual funds or exchange-traded funds that provide investors with an asset allocation aimed at fulfilling a target (goal) by a specified target date (e.g. a retirement date). Typically, as the target date approaches, the asset allocation shifts towards a more conservative, i.e., less risky strategy. During earlier time periods, there is a greater emphasis on investments in equities, while later time periods are characterized by a higher concentration of investments in bonds. This pattern is typically illustrated by glide paths \cite{blanchett} - functions that show the percentage of wealth invested in a certain type of asset over time. However, using fixed glide paths can be suboptimal; for example, in situations when the target date draws close and the goal is not yet accomplished, it is clear that more risk should be taken on through larger positions in equity. This begets a new risk paradigm  \cite{capponi} according to which risk is not directly associated with the volatility of underlying assets, as is the case in traditional portfolio optimization, but rather with the prospect of not achieving the investment goal. Since GBWM involves making a series of investment decisions over time in fluctuating market conditions, each affecting the future position of the investor, it is naturally framed as a problem in sequential decision-making under uncertainty. Reinforcement learning (RL), due to its capacity to tackle sequential decision-making tasks in a data-driven fashion, offers a particularly promising path to GBWM, especially through its model-free and deep (DRL) algorithms. Multiple applications of (D)RL in quantitative finance exist, ranging from standard portfolio optimization ~\cite{zhang, theate}, to market making\footnote{Unlike in GBWM, in market making, increasing levels of risk are typically incurred as the terminal time is approached \cite{avellaneda}, resulting in weaker inventory penalization.} \cite{spooner, gasperov, gasperov2} and optimal trade execution ~\cite{lin}. On the other hand, applications of (D)RL to GBWM are still very scarce, despite its vast potential for the field. In this paper, a novel approach for GBWM based on DRL, with a focus on the robustness of the resulting strategies, 
is proposed. We demonstrate its superior performance over several standard GBWM benchmarks on both simulated and historical market data.

\section{Related Work}  \label{section_2}

%\subsection{Analytical approaches}
%A plethora of standard, analytical approaches to GBWM \cite{forsyth2,merton,das1} exist. Some scholars focus exclusively on retirement investing (\cite{gomes}) and hence use the term "life cycle portfolio choice" instead, while some refer to GBWM as the allocation of target date funds (\cite{roncalli}). Regardless of the used term, the vast majority of proposed methods lead to time-dependent risk mitigation. 

\subsection{Deterministic Glide Path}
A deterministic glide path is a simple approach for GBWM. While essentially a heuristic, it has been adopted by many investors due to its simplicity and intuitive appeal. Note that rules of thumb, such as \textit{100-age}, which suggests that an individual's stock allocation should be equal to $100$ minus their age, are frequently used in retirement asset allocation \cite{hickman1, bodie1}. However, this strategy has been criticized for its sole reliance on the time remaining to the target date \cite{forsyth1, das1}. Despite this, deterministic glide paths are frequently used for target date fund allocation as they provide a straightforward and systematic approach to managing risk as the target date approaches \footnote{An example is given by the asset allocation of Fidelity Freedom Funds (\url{https://www.fidelity.com/mutual-funds/fidelity-fund-portfolios/freedom-funds}).}. We focus on the form:
$$
    \alpha_t = 1- \frac{t}{T},
$$
where $\alpha_t$ is the portfolio weight of the stock at time $t$ and $T$ is the target time.

\subsection{Merton's Constant} 
In Merton's seminal work on lifetime portfolio selection \cite{merton}, it is assumed that the riskless asset has a constant rate of return $r$, while the price of the risky asset $(S_t)_t$ follows the dynamics
$ dS_t = \mu S_t dt + \sigma S_t dZ_t. $
Here, $Z(t)$ is a standard Brownian motion, $\mu$ is the expected rate of return and $\sigma$ is the volatility of the underlying asset. Under this framework, it is demonstrated that, for an investor with a Constant Relative Risk Aversion (CRRA) utility\footnote{The CRRA utility $\mathcal{U}$ is given by: 
$
   \mathcal{U}(x) =  {x^{\gamma}}/{\gamma},\text{ }\gamma < 1,
$
where $1-\gamma$ is the coefficient of relative risk aversion.}, it is optimal to maintain constant portfolio weights of each asset. The optimal weight of the risky asset equals:
$$ \alpha_t =  \frac{\mu - r}{ (1-\gamma) \sigma^2}.$$

%mozda izbaciti roncallija skroz:
\subsection{Variance Budgeting} 
Bruder \emph{et al.} \cite{roncalli} describe an individual's risk aversion by specifying the maximum cumulative portfolio variance they are willing to take on over the investment period, called the variance budget.
%Bruder, Culerier, and Roncalli \cite{roncalli} suggest limiting the cumulated variance of the portfolio instead of using a utility function to describe one’s risk aversion. The variance restraint represents the maximum amount of risk that the investor is willing to take over the investment period. %Bruder \emph{et al.} 
The authors consider this approach (restricted to the universe of two assets) and model their objective as maximizing the expected wealth at maturity $ \alpha_t = \argmax \mathbb{E} [X_T] $ subject to a predefined amount of risk
$ \int_{0}^{T} \alpha_t^2 \sigma_t^2 X_t^2 dt \leq V^2,$
where $V^2$ is the total variance budget of the strategy from the start date to the target date. The optimal allocation strategy is given by:
$$
    \alpha_t = \frac{V}{\sigma_t \sqrt{T} X_t}.
$$

\subsection{Dynamic Programming}
\begin{figure}
 \centering
 \includegraphics[scale=0.5]{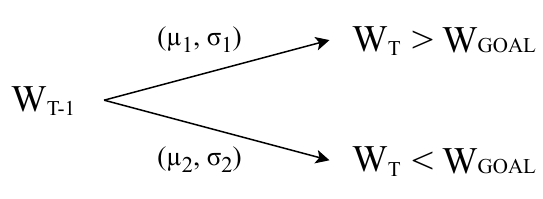}
 \caption{Simplified diagram of the dynamic programming approach} \label{fig4}
 \end{figure}

Das et. al \cite{das1} developed a discrete-time dynamic programming algorithm to create a portfolio trading strategy that maximizes the probability of an investor reaching its target wealth within a predetermined time frame. The approach utilizes portfolios from the efficient frontier, selecting one efficient portfolio at time step $t$ to hold until the next period $t+1$. A state space consisting of time and wealth is divided into individual states by a grid. States are evaluated based on how likely they are to lead to states in which the goal is reached. This evaluation is used to determine, at each time step, the portfolio that is most likely to lead to states with larger values. Figure \ref{fig4} is presented for a better understanding of the method. At time $T-1$, the investor's wealth is $W_{T-1}$. The investor then chooses the portfolio pair $(\mu, \sigma)$ from the efficient frontier that maximizes the probability of achieving the goal wealth at time $T$. The value of the state $W_{T-1}$ is obtained and is then used to evaluate earlier states using backward recursion.
In order to compute these probabilities, the evolution of the portfolio wealth needs to be modeled. The authors choose the Geometric Brownian motion, remarking that the method is also applicable to other models. 
%($dW_t = \mu W_t dt + \sigma W_t dZ_t$)

\subsection{Reinforcement Learning-Based Approaches}
Pendharkar and Cusatis \cite{pendharkar} use RL to construct a two-asset personal retirement portfolio optimized for long-term performance (i.e., a period of a decade or more). Yet the proposed approach uses neither goal-based reward criteria nor time as a state variable, making it difficult to classify it under GBWM. Dixon and Halperin \cite{dixon} propose two practical algorithms - G-Learner, a generative framework, and GIRL, an extension of G-Learning to the setting of inverse RL. The focus is on their use for GBWM (optimization of retirement plans) in the context of robo-advising services. Another RL-based approach is provided by Das and Varma \cite{das2}, where Q-learning is used to obtain the same strategy as in their previous work using dynamic programming \cite{das1}. The comparative advantages of the RL approach, particularly its superior handling of larger state and action spaces, are accentuated by the authors.

\section{Methodology}
We consider the GBWM framework as introduced in \cite{das1, das2} and significantly expand it, primarily by a) putting forth a non-tabular approach based on deep neural networks (NNs) and b) introducing more rigorous training and testing procedures with an emphasis on robustness and generalization. 

\subsection{Markov Decision Process for Goal-Based Wealth Management}
The underlying problem is modeled as a discrete-time Markov Decision Process (MDP) and then approached as an episodic RL task. An MDP is defined as a quintuple $ (S,A,P,R, \gamma )$, where $S$ is the state space, $A$ the action space, $P : S \times A \times S \mapsto [0,1]$ a transition probability function, $R: S \times A \mapsto R$ a reward function and $\gamma \in [0,1]$ a discount factor. 

\subsubsection{State Space}
The state at time $t$ is defined as
$s_t = ( \frac{t}{T}, \frac{W_t}{W_G} )$, where $W_t$ is the current total wealth, $T$ the target date, %(the time when the goal is to be reached, also called the investment horizon), 
and $W_G$ the goal wealth. $W_t = w_B B_t + w_S S_t$, where $B_t$ ($S_t$) is the price of the riskless (risky) asset at time t, which fluctuates in time, and $w_B$ ($w_S$) the corresponding amounts held by the investor. We assume $w_B, w_S \geq 0$, i.e., short (negative) positions are not allowed.
Clearly, $0 \leq \frac{t}{T} \leq 1$ and $0 \leq \frac{W_t}{W_G} < \infty$ (since $B_t \geq 0, S_t \geq 0$). The inclusion of $\frac{W_t}{W_G}$, a simple measure of how close the investor is (in relative terms) to achieving their goals, induces position-dependent behavior in policies, which is in line with the GBWM risk paradigm. The state space variables are normalized with respect to the reference (target) values.

\subsubsection{Action Space}
The action at time $t$ is given by:
$a_t = w_S$, where $w_S \in [0, 1] $ represents the proportion of funds invested in the risky asset $S_t$ at time t. Since $w_S + w_B = 1$, the weight corresponding to the riskless asset is uniquely defined. %The action-space is hence continuous and one-dimensional.

\subsubsection{Reward Function}
GBWM objectives are naturally framed as binary goals. Consequently, the RL reward is only received at the end of the episode, provided the goal is reached ($r_T = \mathbbm{1}_{ \{ W_T\geq W_G \}}$). Otherwise, the rewards equal zero. This is a case of sparse rewards, which generally tend to be more difficult to learn from.

\subsection{Dataset}
Our study uses a two asset model, which is a common approach in goal-based investing to balance capital preservation and growth. While this approach somewhat limits the diversity of the portfolio, this paper focuses on a simple model which allows a direct comparison with other common approaches in the literature. We employ a dataset of monthly returns of the S\&P 500 index as the risky asset and bonds as the risk-free asset, made available by R. Shiller\footnote{\url{http://www.econ.yale.edu/~shiller/data.htm}}. The dataset was split into a training set covering returns from 1901 to 1991 and a testing set covering returns from 1992 to 2022.

\subsection{Training Procedure}
First, each episode, representing $10$ trading years, is split into $120$ time steps of equal length, each corresponding to a single trading month. At the beginning of each time step, the investor decides what percentage to invest in the (non-)risky asset and allocates the wealth accordingly. This procedure is iterated until the target date is reached and the episode terminates. Considering that historical data only provides a single trajectory, while simulating data from a multivariate normal distribution with fixed parameters may detract from realism, we propose an alternative data simulation mechanism, presented below. The goal here is to both extract as much as possible from the historical market data and provide the DRL agent with a sufficiently large training set. For simplicity, it is assumed that the investor begins at state $s_0 = (0, 0.6)$, i.e., with the initial investment equal to 60\% of the goal.

\subsubsection{Data Generation Procedure}\label{subsection1}
The training dataset is represented by \sloppy $\{ \bm{R_1}, \ldots, \bm{R_{N_{train}}} \}$, where $\bm{R_t} = (R_{t}^{bond},\ R_{t}^{stock})$ contains bond and stock returns at time $t$. The trajectories used for training the DRL agent are generated by the following procedure:
\begin{enumerate}
    \item choose an index $k \in \{n+1, ..., N_{train}\}$ randomly, 
    \item \label{step2} use the series of returns $\{ \bm{R_{k-n}}, ..., \bm{R_k}\}$ to estimate the mean vector $ \bm{\mu} = (\mu^{bond},\ \mu^{stock})$ and the covariance matrix $\mathbf{\Sigma}$, where $n$ denotes the window size of returns used for estimation ($n < k$),  
    \item sample the multivariate normal distribution $\mathcal{N}(\boldsymbol{\mu}, \mathbf{\Sigma})$ to obtain a trajectory.
\end{enumerate}
The parameter $n$ was set to $120$ months, the same as the trajectory (and episode) length. The main idea behind this type of procedure is to generate a variety of trajectories emanating from different distributions, with the goal of enhancing the robustness of the DRL agent by exposing it to varying conditions during training. 

\subsubsection{Algorithm and Network Architecture}
Tabular RL methods (as seen in \cite{das2}) have difficulties in evaluating the values of rarely encountered states. This makes it challenging for the RL agent to determine the optimal policy. DRL approaches, employing function approximation, hence offer a more promising path. We use a state-of-the-art, actor-critic-based %on-policy 
algorithm Proximal Policy Optimization (PPO) \cite{Schulman}. The objective function used by PPO is given by: % The key idea behind PPO is to optimize a surrogate objective function that approximates the true objective. The surrogate objective function is ... Policy gradient methods are usually very sensitive to the choice of the learning rate. To regulate the policy gradient update and make sure that the new policy will not diverge too far from the old one, PPO's objective function is given by:
\begin{equation*}
J(\boldsymbol{\theta}) = \mathbb{E} \left[  \min \left( r_{\boldsymbol{\theta}} A(s,a), \textrm{ clip} (  r_{\boldsymbol{\theta}}, 1-\epsilon, 1+ \epsilon ) A(s,a) \right)  \right],
\end{equation*}
where $\boldsymbol{\theta}$ denotes the policy parameters, $\mathbb{E}$ the empirical expectation, $r_{\theta}$ the ratio of probabilities under the new and old policies, $\epsilon$ a hyperparameter, and $A$ the advantage function. The clipping factor ($\text{clip}$) is used to prevent large policy changes. %With this objective function, PPO becomes simpler to implement while preserving stability and reliability. 
A feed-forward NN with $2$ hidden layers, $6$ neurons each, is employed, with the ReLU activation function. The discount factor $\gamma$ is set to $1$, and the choice of the very small learning rate of $0.0001$ was guided by the stochasticity of the environment. The Stable Baselines3 \cite{stablebaselines3} library and PyTorch were used for the implementation. 

\section{Results}
\subsection{Reinforcement Learning Policy}
\begin{figure} \centering
 \includegraphics[width=10cm]{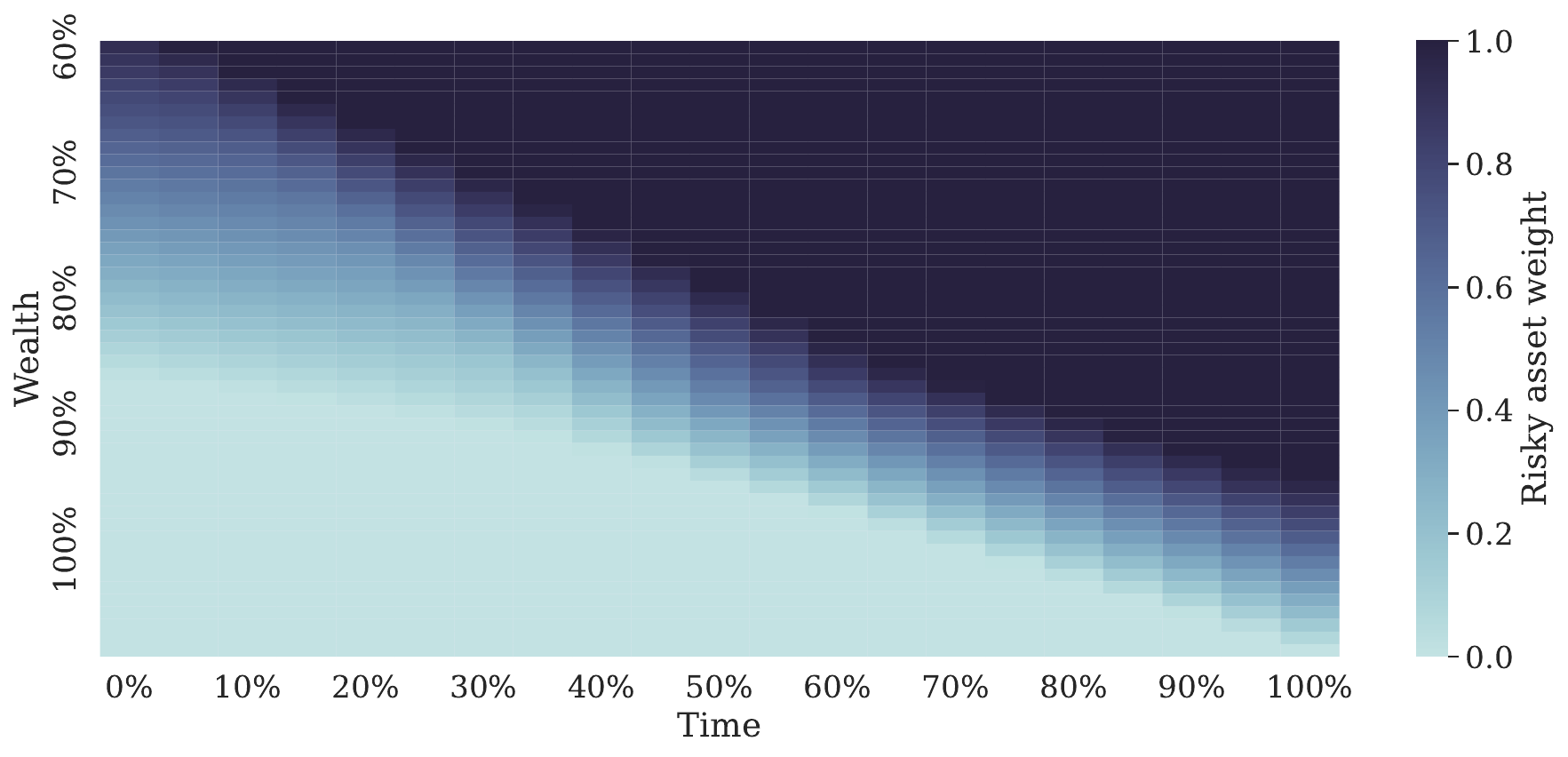}
 \caption{Policy learned by the DRL agent. The agent was trained for $200 000$ episodes, $120$ steps each, and the training wall-clock time was around $1$ hour with the available computational resources.} \label{fig1}
 \end{figure}

The resulting DRL policy is shown\footnote{Since the original policy found by PPO is stochastic, its determinism is enforced by returning the mode of the distribution over the action space instead of sampling from it.} in Fig.~\ref{fig1}.
It depicts the state space and the actions selected by the agent in each state. If the agent is in a certain state, the color displayed on the heatmap presents the agent's action. The policy's dependence on both wealth and time is evident. The obtained policy is intuitive, easily interpretable, and satisfies properties associated with a glide path investment strategy, specifically taking on greater risk earlier on in the investment period. The agent uses a less risky portfolio when the current amount of wealth is sufficient or when there is still enough time to achieve the target. In contrast, a riskier portfolio is favored when the goal has not yet been reached despite the end of the episode drawing near.

\subsection{Performance Results}

\begin{figure}[h!]
    \centering
    \subfloat[\centering Merton's constant ]{{\includegraphics[width=0.4\textwidth]
    {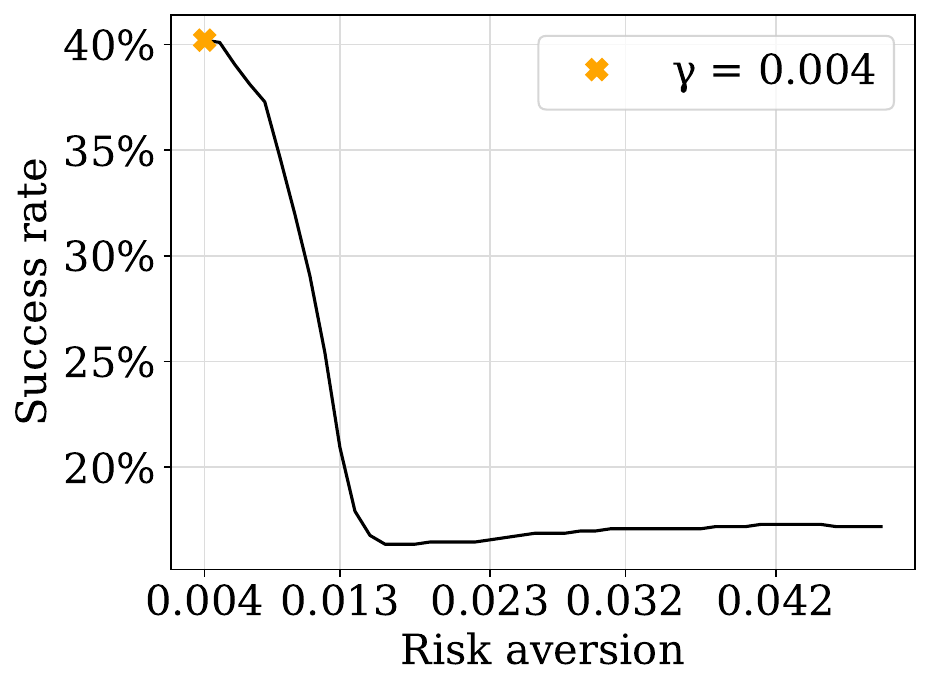} } \label{subfig1}}%
    \qquad
    \subfloat[\centering Variance budgeting]{{\includegraphics[width=0.4\textwidth]{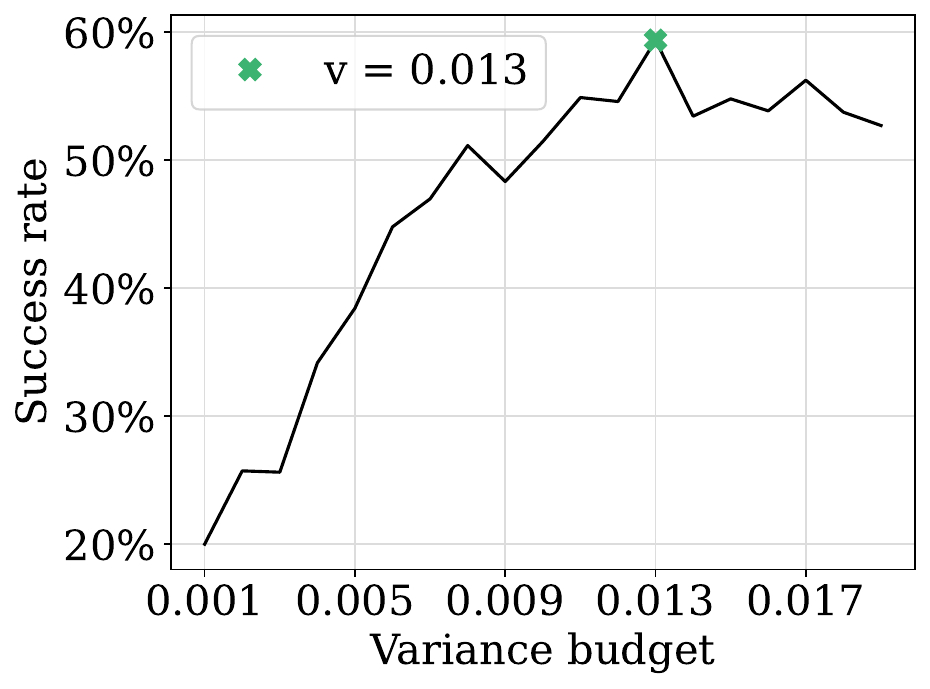} } \label{subfig2}}%
    \caption{Choice of parameters for benchmark methods}%
    \label{fig3}
\end{figure}

%ovdje dodat/prmojiniet za mertona i roncallija
The analytical and numerical approaches presented in Section \ref{section_2} were used as benchmarks to evaluate the performance of the RL agent fairly. This ensures rigorous testing with as many as four benchmarks, surpassing the number of benchmarks used in previous literature. The risk parameters required for determining  Merton's constant (risk aversion parameter $\gamma$) and the variance budgeting approach were chosen based on their performance on the training set. Figure \ref{fig3} displays the percentage of successful episodes per value for each method, i.e., those that reached the goal for a certain fixed parameter. Subfigure \ref{subfig1} shows the selection of the risk aversion parameter for Merton's constant, which was searched for in the interval $[0.004, 0.05]$ to include all possible options for the constant strategy. This interval was selected because $\gamma \leq 0.004$ produces a stock-only portfolio, i.e., $\alpha=1$, while values of $\gamma \geq 0.05$ result in a bond-only portfolio, i.e., $\alpha=0$. The intermediate values of $\gamma$ yield mixed portfolios. The best result on the training set was achieved with $\gamma = 0.004$. For the variance budgeting approach, the optimal risk budget on the training set was $v=1.3\%$, as depicted in Subfigure \ref{subfig2}. The parameter was searched within the range of $[0.001, 0.02]$ for the same reason as in the case of Merton's constant. 
%The optimal variance budget is much smaller than in the original paper where this method was proposed due to the use of monthly variance which is much smaller than the yearly variance used in the original paper. 

%\vspace{-3cm}
To demonstrate the robustness of the trained RL agent, three distinct methods of testing were employed, based on the use of a) real historical trajectories, b) simulated data, and c) bootstrapped data.%, which will be explained in the following subsections. 
The aggregated results are shown in Table \ref{tab1}, with the following abbreviations:
DG -- Deterministic glide path, MC -- Merton's constant, VB -- Variance budgeting, DP -- Dynamic programming, RL -- Reinforcement learning.

\begin{table}[h!]
\setlength{\tabcolsep}{5.5pt}
\renewcommand{\arraystretch}{1.2}
\centering
\caption{Comparison of results - DRL vs benchmarks}\label{tab1}
\begin{tabular}{l|c|ccc|ccc}
\hline \hline
                         & \multicolumn{1}{l|}{Historical data} & \multicolumn{3}{c|}{Simulated data}        & \multicolumn{3}{c}{Bootstrapped data}    \\ 

&  & \multicolumn{3}{c|}{(window size)}        & \multicolumn{3}{c}{(block size)}    \\
                         
                         %\hdashline
\multicolumn{1}{c|}{}    & \multicolumn{1}{c|}{}         & 36               & \{24, 36, 48\} & 60               & 1                & \{1,2,3\}        & \{4,5,6\}        \\ \hline
DG & 54.3\%                       & 70.5\%          & 70.3\%            & 69.7\%          & 79.2\%          & 76.9\%          & 75.1\%          \\
MC        & 67.0\%                       & 71.8\%          & 72.3\%            & 67.0\%          & 80.7\%          & 77.4\%          & 74.6\%          \\
VB      & 68.6\%                       & 74.1\%          & 74.1\%            & 70.6\%          & 87.2\%          & 83.5\%          & 80.1\%          \\
DP      & 74.4\%                       & 73.7\%          & 73.8\%            & 74.7\%          & 88.9\%          & 84.4\%          & 80.7\%          \\
RL   & \textbf{77.5\%}              & \textbf{76.7\%} & \textbf{76.4\%}   & \textbf{75.3\%} & \textbf{90.8\%} & \textbf{86.3\%} & \textbf{82.3\%} \\ \hline \hline
\end{tabular}
\end{table}

\subsubsection{Historical Market Data}
The testing on historical market data was performed by using overlapping historical trajectories of monthly returns from January 1991 to June 2022. The testing dataset includes $378$ data points. Considering that series of lengths $120$ are needed (the number of months in $10$ years), it is possible to generate $258$ different (but overlapping) trajectories from the dataset. Table \ref{tab1} shows the percentages of achieved targets for those time series. It is clear that the DRL agent outperforms all other benchmarks on historical market data. However, considering both the dependence between individual paths and the limited amount of historical data, it is essential to have another approach to test the robustness and efficacy of the obtained DRL agent. 

\subsubsection{Simulated Data}
\begin{figure} \centering
 \includegraphics[width=0.5\textwidth]{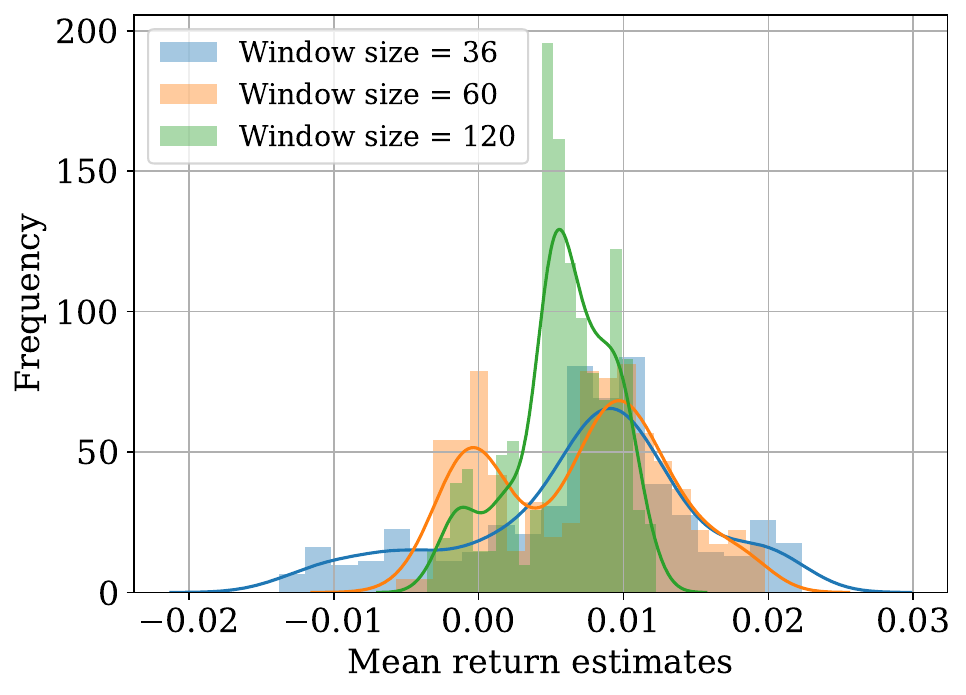}
 \caption{Distribution of mean return estimates for each window size, together with Gaussian kernel density estimates.} \label{fig5}
 \end{figure}

\begin{figure}
 \centering
 \includegraphics[width=0.5\textwidth]{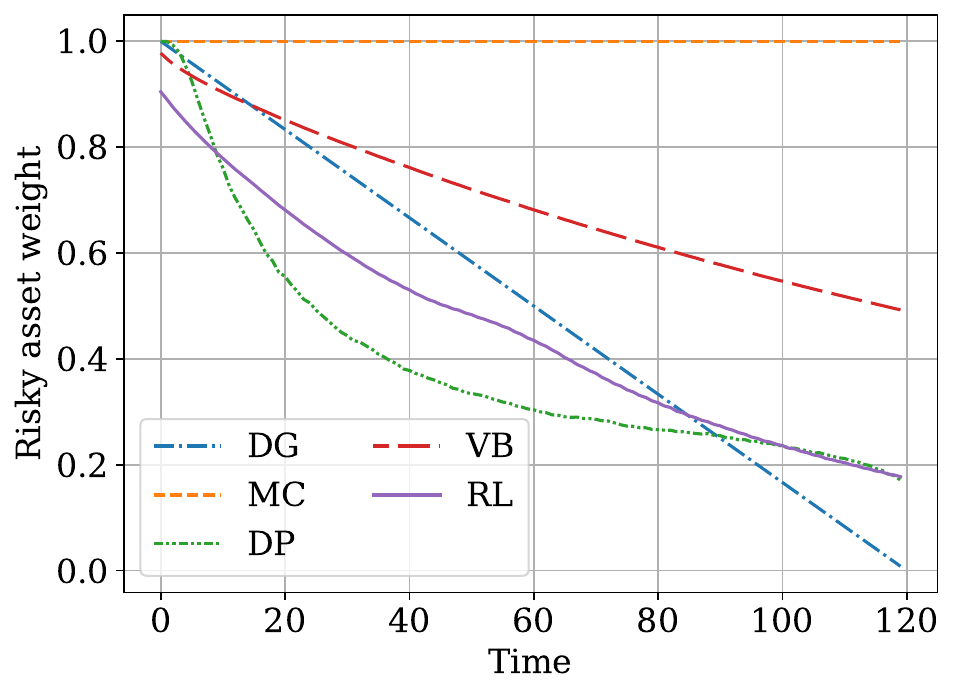}
 \caption{Glide paths of different strategies} \label{fig2}
 \end{figure}

The data was acquired using the training procedure outlined in Subsection \ref{subsection1}, with the only difference being the use of historical market data from 1991 to 2022 as the underlying dataset. To construct a testing set that encompasses a wider range of potential scenarios, simulations were performed with different window sizes for parameter estimation in Step 2 of the procedure. Figure \ref{fig5} shows the distributions of mean return estimates for different window sizes. We note that the use of shorter estimation windows results in an increase in the mean return variance. Consequently, returns generated from distributions calibrated on smaller window sizes are more versatile, leading to more diverse trajectories and, in turn, the enhanced robustness of the RL agent.
Two fixed window sizes, $36$ and $60$, were considered, and an additional experiment was conducted in which a window size was randomly chosen from the set $\{24, 36, 48\}$ for every trajectory anew. For testing purposes, $10000$ trajectories were simulated for each setting. The Table shows the percentage of successful portfolio allocations, i.e., ones that led to achieving the goal. Again, the DRL model outperforms the benchmark methods. Figure \ref{fig2} displays the average proportion of the portfolio invested in the risky asset in the case of the window length equal to $36$. Each point on the graph represents the average allocation of a given method at a specific time step. All of the obtained glide paths with the sole exception of MC indicate risk reduction over time, as is typical for GBWM.
\subsubsection{Bootstrapped Data} 
The bootstrap, originally presented by Efron \cite{efron}, is a statistical method that involves generating new data by simulating from an existing data set. It is a resampling approach in which multiple new samples are generated from the original dataset with replacement, thereby creating a new dataset that captures the variability and uncertainty of the original data. This technique was used to generate $10000$ test trajectories from historical returns. Additionally, a variation of the bootstrap method known as block bootstrapping was employed. Block bootstrapping is specifically designed for time series data and involves sampling blocks of consecutive observations, preserving the correlations within each block while still generating synthetic data. It was used to generate $10000$ trajectories, each one generated by first choosing a block size $b$ from a given set of values, and then sampling from the testing set to acquire a $120$-month-long series. If $b=1$ the method deteriorates to regular bootstrapping. For $b=2$ a trajectory is obtained by sampling $60$ blocks of consecutive returns of two months. Three variations of block sizes were used for testing, and the corresponding success rates are presented in Table \ref{tab1}. 
\\~\\ 
\indent Given that the agent was trained on data generated by the simulator, it has been exposed to a much wider range of market conditions than those in the historical test set. However, all of the considered methods (all the benchmark methods and the proposed approach) seem to perform better on out-of-sample simulated data than on historical market data. This points to the fact that historical market data likely exhibits more complex dynamics which cannot easily be replicated by the simulation model — however, the simulation model allows us to generate a large number of trajectories, which is crucial in training the agent. Moreover, the results evidently testify to the fact that the proposed approach yields improved performance over the benchmarks in all of the considered test cases, including historical data.

\section{Conclusion and Future Work}

In this paper, a novel approach for goal-based wealth management based on deep reinforcement learning is presented. The results demonstrate that the proposed method outperforms multiple established benchmarks on both historical and simulated market data, using multiple testing procedures. Our study provides evidence that the proposed approach can be a valuable addition to existing wealth management strategies, as it offers improved performance and potential for practical applications in various financial settings. Despite the challenges posed by the complexity of decision-making processes in deep reinforcement learning, this paper presents a highly explainable policy for the agent, indicating that progress is being made in improving the interpretability of these systems. Future work should consider the following possibilities: First, the present work could be used in the context of regime-based asset allocation~\cite{nystrup}, for example by expanding the state space to include market regime-based features. This would require the development of a more complex financial market simulator capable of modeling non-stationary effects. Second, the approach might be recast from the perspective of more sophisticated GBWM frameworks that take into account multiple future goals \cite{das1, capponi2}. Third, different risk preferences might be considered to pave the way toward more personalized wealth management. Using a non-binary reward function, as opposed to solely aiming to attain the predetermined target wealth, is expected to lead to more precise catering to investors' preferences. Additionally, cash infusion during the investment period could also be incorporated into the framework, as is typical in retirement and goal-based investing \cite{das1}. Lastly, generalizations to multi-asset scenarios present a potentially fruitful further step, as multi-asset allocation provides a more diversified investment portfolio, reducing overall risk and increasing the likelihood of achieving financial goals.

\subsubsection{Acknowledgements} This work was supported in part by the Croatian Science Foundation under Project 5241, and in part by the European Regional Development Fund under Grant KK.01.1.1.01.0009 (DATACROSS).

\bibliographystyle{splncs04}

\end{document}